\documentstyle[preprint,aps,eqsecnum,epsfig]{revtex} 
\tighten
\begin{document}
\draft
\title{\bf  FRACTAL STRUCTURES AND SCALING LAWS IN THE
UNIVERSE. STATISTICAL MECHANICS OF THE SELF-GRAVITATING GAS}
\author{{\bf  H. J. de Vega$^{(a)}$,
  N. S\'anchez$^{(b)}$ and  F. Combes$^{(b)}$}\bigskip}

\bigskip

\address
{ (a)  Laboratoire de Physique Th\'eorique et Hautes Energies,
Universit\'e Paris VI, Tour 16, 1er \'etage, 4, Place Jussieu
75252 Paris, Cedex 05, FRANCE. Laboratoire Associ\'e au CNRS UMR 7589.\\
(b) Observatoire de Paris,  Demirm, 61, Avenue de l'Observatoire,
75014 Paris,  FRANCE. 
Laboratoire Associ\'e au CNRS UA 336, Observatoire de Paris et
\'Ecole Normale Sup\'erieure.  \\ }

\date{June 1998}
\maketitle
\begin{abstract}
Fractal structures are observed in the universe in two very different ways.
Firstly, in the gas forming the cold interstellar medium in  scales 
from $ 10^{-4}$pc till $100$pc. Secondly, the galaxy distribution has
been observed to be  fractal in scales up to hundreds of Mpc. 
We give here a short review of the statistical mechanical 
(and field theoretical) approach developed by us
for the cold interstellar medium (ISM) and large structure of
the universe.
We consider a non-relativistic self-gravitating gas in thermal equilibrium
at temperature $T$ inside a volume $V$.
The statistical  mechanics of such system has  special features and,
as is known,  the 
thermodynamical limit does not exist in its customary form.
Moreover, the treatments through microcanonical, canonical and grand
canonical ensembles yield different results.
We present here for the first time the equation of state for the
self-gravitating gas in the canonical ensemble. We find that it has the form
$ p = [N T/ V] f(\eta) $, where $ p $ is the pressure, 
$N$ is the number of particles and 
 $ \eta \equiv {G \, m^2 N \over V^{1/3} \; T}$. The 
$ N\to \infty $ and $ V \to \infty $ limit exists keeping $ \eta $ fixed.
We compute the function  $ f(\eta) $ using Monte Carlo simulations 
and for small $\eta $, analytically. We compute the thermodynamic
quantities of the system
as  free energy, entropy, chemical potential, specific heat,
compressibility and speed of sound. We reproduce the well-known
gravitational  phase transition associated to 
the Jeans' instability. Namely, a gaseous phase for $ \eta < \eta_c $
and a condensed 
phase for  $ \eta > \eta_c $.
Moreover, we derive the precise behaviour of the physical
quantities near the transition. In particular, the pressure vanishes as
$ p \sim(\eta_c-\eta)^B $ with $ B \sim 0.2 $ and $ \eta_c \sim 1.6 $
and the energy fluctuations diverge as 
$ \sim(\eta_c-\eta)^{B-1} $. The speed of sound decreases monotonically  with  
$ \eta $ and approaches the value $ \sqrt{T/6} $ at the transition.
\end{abstract}
\section{Statistical Mechanics of  the Self-Gravitating  Gas}

Physical systems at thermal equilibrium are usually homogeneous. This is the 
case for gases with 
short range  intermolecular forces (and in absence of external fields).
When long range interactions as the gravitational force are present even
the ground state is often  inhomogeneous. In this case,  each element of the
substance is acted on by very strong forces due to distant
particles of the gas. Hence regions near to and far from the boundary of the 
volume occupied by the gas will be in very different conditions and as a 
result 
the homogeneity of the gas is destroyed \cite{llms}. This  basic  inhomogeneity
suggests that fractal structures can arise in a self-interacting gravitational 
gas\cite{natu,prd,gal,eri,pcm}.

Let us review very briefly our recent work \cite{natu,prd,gal,eri} on the
statistical properties of 
a self-interacting gravitational gas in thermal equilibrium. We
discussed two relevant astrophysical applications of such gas: the cold 
interstellar medium (ISM) and the galaxy distributions.

In the grand canonical ensemble, we showed that the
self-gravitating gas is exactly equivalent to a field theory 
of a single scalar field $ \phi({\vec x}) $ with exponential self-interaction.
We analyzed this field theory perturbatively and non-perturbatively
through the renormalization group approach. We showed {\bf scaling}  
behaviour (critical) for a continuous range of the temperature and of 
the other physical parameters. We derive in this framework the scaling
relation  
$$ 
M(R) \sim R^{d_H} 
$$
 for the mass on a region of size $ R $,
and 
$$
 \Delta v \sim R^q 
$$ 
for the velocity dispersion where 
$ q = \frac12(d_H -1) $. For the density-density correlations
we find a power-law  behaviour for large distances 
$$ 
\sim |{\vec r_1} -{\vec r_2}|^{2 d_H -6} \; .
$$
  The fractal dimension
$  d_H $ turns to be related with the critical exponent $ \nu $ of the 
correlation length by 
$$  
d_H = 1/ \nu \; .  
$$
Mean field theory yields for the scaling
exponents $ \nu = 1/2  , \; d_H = 2 $ and $ q = 1/2 $. Such values
are compatible with the present ISM observational data: $  1.4    \leq
d_H    \leq   2     ,   \; 0.3  \leq     q  \leq 0.6 \;  $. 

We developed in ref.\cite{gal} a field theoretical approach to the
galaxy distribution. We 
consider a gas of self-gravitating masses on the
Friedman-Robertson-Walker background, 
 in quasi-thermal equilibrium. We derive the galaxy
correlations using renormalization group methods.  We find that the
connected $N$-points density  correlator 
$ C({\vec r}_1,{\vec r}_2,\ldots,{\vec r}_N) $ scales as  
$$ 
r_1^{N(D-3)} \; , 
$$
when $ r_1 >> r_i, \; 2\leq i \leq N $. 
There are no free parameters in this theory.

Our  study of the statistical mechanics of a self-gravitating system indicates 
that gravity provides a dynamical mechanism to produce   fractal
structure.

This paper is organized as follows. In section II we summarize the 
main properties of  the ISM, in section III we review the relevant
aspects of the large scale structure of the universe, in sec. IV
we develop the statistical mechanics of the self-gravitating gas in
the canonical ensemble.  Discussion and remarks are presented in section V. 

\section{The Interstellar Medium}

The interstellar medium (ISM) is a gas essentially formed by atomic (HI) 
and molecular ($H_2$) hydrogen, distributed in cold ($T \sim 5-50 K$) 
clouds, in a very inhomogeneous and fragmented structure. 
These clouds are confined in the galactic plane 
and in particular along the spiral arms. They are distributed in 
a hierarchy of structures, of observed masses from 
$10^{-2} \; M_{\odot}$ to $10^6 M_{\odot}$. The morphology and
kinematics of these structures are traced by radio astronomical 
observations of the HI hyperfine line at the wavelength of 21cm, and of
the rotational lines of the CO molecule (the fundamental line being
at 2.6mm in wavelength), and many other less abundant molecules.
  Structures have been measured directly in emission from
0.01pc to 100pc, and there is some evidence in VLBI (very long based 
interferometry) HI absorption of structures as low as $10^{-4}\; pc = 20$ AU 
(3 $10^{14}\; cm$). The mean density of structures is roughly inversely
proportional to their sizes, and vary 
between $10$ and $10^{5} \; atoms/cm^3$ (significantly above the 
mean density of the ISM which is about 
$0.1 \; atoms/cm^3$ or $1.6 \; 10^{-25}\; g/cm^3$ ).
Observations of the ISM revealed remarkable relations between the mass, 
the radius and velocity dispersion of the various regions, as first 
noticed by Larson \cite{larson}, and  since then confirmed by many other 
independent observations (see for example ref.\cite{obser}). 
From a compilation of well established samples of data for many different  
types of molecular clouds of maximum linear dimension (size) $ R $,  
total mass $M$ and internal velocity dispersion $ \Delta v$ in each region: 
\begin{equation}\label{vobser}
M (R)  \sim    R^{d_H}     \quad        ,     \quad  \Delta v \sim R^q \; ,
\end{equation}
over a large range of cloud sizes, with   $ 10^{-4}\; - \; 10^{-2}
\;  pc \;   \leq     R   \leq 100\;  pc, \;$
\begin{equation}\label{expos}
1.4    \leq   d_H    \leq   2     ,   \;     0.3  \leq     q  \leq
0.6 \; . 
\end{equation}
These {\bf scaling}  relations indicate a hierarchical structure for the 
molecular clouds which is independent of the scale over the above 
cited range; above $100$ pc in size, corresponding to giant molecular clouds,
larger structures will be destroyed by galactic shear.

These relations appear to be {\bf universal}, the exponents 
$d_H , \; q$ are almost constant over all scales of the Galaxy, and
whatever be  
the observed molecule or element. These properties of interstellar cold 
gas are supported first at all from observations (and for many different 
tracers of cloud structures: dark globules using $^{13}$CO, since the
more abundant isotopic species $^{12}$CO is highly optically thick, 
dark cloud cores using $HCN$ or $CS$ as density tracers,
 giant molecular clouds using $^{12}$CO, HI to trace more diffuse gas, 
and even cold dust emission in the far-infrared).
Nearby molecular clouds are observed to be fragmented and 
self-similar in projection over a range of scales and densities of 
at least $10^4$, and perhaps up to $10^6$.

The physical origin as well as the interpretation of the scaling relations 
 (\ref{vobser}) are not theoretically understood. 
The theoretical derivation of these
 relations has been the subject of many proposals and controversial 
discussions. It is not our aim here to account for all the proposed models 
of the ISM and we refer the reader to refs.\cite{obser} for a review.

The physics of the ISM is complex, especially when we consider the violent
perturbations brought by star formation. Energy is then poured into 
the ISM either mechanically through supernovae explosions, stellar winds,
bipolar gas flows, etc.. or radiatively through star light, heating or
ionising the medium, directly or through heated dust. Relative velocities
between the various fragments of the ISM exceed their internal thermal
speeds, shock fronts develop and are highly dissipative; radiative cooling
is very efficient, so that globally the ISM might be considered 
isothermal on large-scales. 
Whatever the diversity of the processes, the universality of the
scaling relations suggests a common mechanism underlying the physics.

  We propose that self-gravity is the main force at the origin of the 
structures, that can be perturbed locally by heating sources. 
Observations are compatible with virialised structures at all scales.
 Moreover, it has been suggested that the molecular clouds ensemble is
in isothermal equilibrium with the cosmic background radiation at $T \sim 3 K$
in the outer parts of galaxies, devoid of any star and heating
sources \cite{pcm}. This colder isothermal medium might represent the ideal
frame to understand the role of self-gravity in shaping the hierarchical
structures. Our aim is to show that the scaling laws obtained are then
quite stable to perturbations.

Till now, no theoretical derivation of the scaling laws 
eq.(\ref{vobser}) has been 
provided in which the values of the exponents are {\bf obtained} from the 
theory (and not just taken from outside or as a starting input or hypothesis).

The aim of our work \cite{natu,prd,gal,eri} 
 is to develop a theory of the cold ISM. A first 
step in this goal was to provide a theoretical derivation of the scaling 
laws eq.(\ref{vobser}), in which the values of the exponents $d_H , \; q$ are
{\bf obtained} from the theory\cite{natu,prd,gal,eri}.
For this purpose, we  implemented for the ISM the powerful tool of field 
theory and the Wilson's approach to critical phenomena \cite{rg}.

\section{Galaxy Distributions}

One obvious feature of galaxy and cluster distributions in the sky is their
hierarchical property: galaxies gather in groups, that are embedded in
clusters, then in superclusters, and so on\cite{sha,abe}.

The knowledge of the galaxy and cluster correlations allows a more precise
characterization of their distributions. Unfortunately, the most  widely spread
two point correlation function $\xi(r)$ in galaxy distributions
studies, is based on the assumption that the 
Universe reaches homogeneity on a scale smaller than the sample size.
 $\xi(r)$ is  defined as
$$
\xi(r) = \frac{<n(r_i).n(r_i+r)>}{<n>^2} -1
$$
where $n(r)$ is the number density of galaxies, and $<...>$ is the volume
average (over $d^3r_i$). The  length $r_0$ is defined
by $\xi(r_0) = 1$. The function $\xi(r)$ has a power-law behaviour 
of slope $-\gamma$ for $r< r_0$, then it turns down to zero 
rather quickly at the statitistical limit of the sample. This rapid
fall leads to an over-estimate of the small-scale $\gamma$. One finds
the slope $\gamma$, the same for galaxies and clusters, of $\approx$
1.7  (e.g. \cite{peeb}).

It has been shown in refs.\cite{copies} and \cite{colpie} that the homogeneity 
hypothesis could perturb significantly the results.

Pietronero \cite{pietro} introduces the conditional density
$$
\Gamma(r) = \frac{<n(r_i).n(r_i+r)>}{<n>} 
$$
which is the average density around an occupied point.
For a fractal medium, where the mass depends on the size as
$$
M(r) \propto r^D
$$ 
$D$ being the fractal (Haussdorf) dimension, the conditional
density behaves as
$$
\Gamma(r) \propto r^{D-3}
$$
This is exactly the statistical analysis used for the interstellar
clouds, since the ISM astronomers have not  adopted from the start 
any large-scale homogeneity assumption (cf. \cite{pc}).

The fact that for a fractal the correlation $\xi(r)$ can be highly
misleading is readily seen since
$$
\xi(r) = \frac{\Gamma(r)}{<n>} -1
$$
and for a fractal structure the average density of the sample $<n>$ is a 
decreasing function of the sample length scale. In the general use of
$\xi(r)$,  $<n>$ is taken for a constant, and we can see that
$$
D = 3 - \gamma \quad .
$$
If for very small scales,
both $\xi(r)$ and $\Gamma(r)$ have the same power-law behaviour, with the 
same slope $-\gamma$, then the slope appears to steepen for $\xi(r)$
when approaching the  length $r_0$. This explains why
with a correct statistical analysis \cite{revsyllabpie}, the actual
$\gamma \approx 1-1.5$ is smaller  
than that obtained using $\xi(r)$. This also explains why the amplitude of
$\xi(r)$ and $r_0$ increases with the sample size, and for clusters as well. 

This scale-invariance has suggested very
early the idea of fractal models for the clustering hierachy of galaxies
\cite{deVau,fractal}. Since then, many authors have
shown that a fractal distribution indeed reproduces quite well the 
aspect of galaxy catalogs, for example by simulating a fractal and observing
it, as with a telescope \cite{sss,spee}.

There is some ambiguity in the definition of the two-point correlation
function $\xi(r)$ above, since it depends on the assumed scale
beyond which the universe is homogeneous; indeed it includes a normalisation by
the average density of the universe, which, if the homogeneity scale is not 
reached, depends on the size of the galaxy sample.
Once $\xi(r)$ is defined, one can always determine a  length $r_0$ where 
$\xi(r_0)$ =1 \cite{dp83}.  For galaxies,
the most frequently reported value is $r_0 \approx 5 h^{-1}$ Mpc
(where $h = H_0$/100km s$^{-1}$Mpc$^{-1}$), but it has been shown to increase 
with the distance limits of galaxy catalogs \cite{detal},
$r_0$ is called `correlation length' in the galaxy literature. 
[The notion of correlation length $\xi_0$ is usually different in physics,
where  $\xi_0$ characterizes the exponential decay of correlations $ (\sim 
e^{- r/ \xi_0} ) $. For power decaying correlations, it is said that the  
correlation length is infinite].

The same problem occurs for the two-point correlation function of
galaxy clusters; the corresponding $\xi(r)$ has the same power law 
as galaxies, their  length  $r_0$ has been reported to be about 
$r_0 \approx 25 h^{-1}$ Mpc, and their correlation amplitude is therefore
about 15 times higher than that of galaxies \cite{pgh}.
The latter is difficult to understand, unless there is a considerable
difference between galaxies belonging to clusters and field galaxies (or
morphological segregation). The other obvious explanation is that
the normalizing average density of the universe was then chosen lower.

This statistical analysis of the galaxy catalogs has been criticized in 
refs.\cite{pietro,eina,colpie}, 
who stress the unconfortable dependence of $\xi(r)$ and of the  length $r_0$
upon the finite size of the catalogs, and on the {\it a priori} assumed 
value of the large-scale homogeneity cut-off.  A way to circumvent these 
problems is to deal instead with the average density as a function of size. 
It has been shown that the galaxy distribution behaves as a pure
self-similar fractal over scales up to $\approx 100 h^{-1}$ Mpc,
the deepest scale to which the data are statistically robust \cite{revsyllabpie}.
This is more consistent with the observation of contrasted large-scale
structures, such as superclusters, large voids or great walls of
galaxies of $\approx 200 h^{-1}$ Mpc \cite{dela}
After a proper statistical analysis of all available
catalogs (CfA, SSRS, IRAS, APM, LEDA, etc.. for galaxies, and Abell and
ACO for clusters) Pietronero et al \cite{pietro,revsyllabpie} state that the 
transition to homogeneity might not yet have been reached up to the deepest
scales probed until now. At best, this point is quite controversial,
and the large-scale homogeneity transition is not yet well known. 

Isotropy and homogeneity are expected at very large scales from the
Cosmological Principle (e.g. \cite{peeb}). However, this does not imply
local or mid-scale homogeneity (e.g. \cite{fractal}, \cite{revsyllabpie}.
a fractal structure can be locally isotropic, but inhomogeneous.
The main observational evidence in favor of the Cosmological Principle
is the remarkable isotropy of the
cosmic background radiation (e.g. \cite{cobe}), that provides information
about the Universe at the matter/radiation decoupling. There must therefore
exist a transition between the small-scale fractality to large-scale
homogeneity. This transition is certainly smooth, and might correspond to the
transition from linear perturbations to the non-linear gravitational collapse 
of structures. The present catalogs do not yet see the transition since
they do not look up sufficiently back in time. It can be noticed that
some recent surveys begin to see a different power-law
behavior at large scales ($\lambda \approx 200-400  h^{-1}$ Mpc,
e.g. \cite{lin}). 

\medskip

There are several approaches to understand non-linear clustering, and therefore
the distribution of galaxies, in an infinite gravitating system. 
Numerical simulations have been widely used, in the hope to trace back from 
the observations the initial mass spectrum of fluctuations, and to test 
postulated cosmologies such as CDM and related variants (cf \cite{ost}). 
[That is numerically solving Newton's equations of motion of
self-gravitating particles]. This approach 
has not yet yielded definite results, especially since the physics of
the multiple-phase universe is not well known. Also numerical limitations
(restricted dynamical range due to the softening and limited volume) have
often masked the expected self-similar behavior.

We presented  in \cite{gal}  a new approach based on field theory and the 
renormalisation group to understand the clustering
behaviour of a self-gravitating expanding universe. We also consider
the thermodynamics properties of the system, assuming quasi-equilibrium
for the range of scales concerned with the non-linear regime and
virialisation. Using statistical field theory,
 the renormalisation group and the finite-size scaling ideas,
we determined the scaling behaviour.
The small-scale fractal universe can be considered critical with large density
fluctuations developing at any scale. We derived
the corresponding critical exponents which yielded the fractal
dimension $D$. It is very close
to those measured on galaxy catalogs through statistical
methods based on the average density as function of size; these methods
reveal in particular a fractal dimension $D \approx 1.5-2$ 
\cite{otra,revsyllabpie}.
This fractal dimension is strikingly close to that observed for
the interstellar medium or ISM (e.g. \cite{larson,obser})
We showed in ref.\cite{gal} that the theoretical framework based on 
self-gravity that we have developped for the ISM \cite{natu,prd} is
also  a dynamical mechanism leading to the  fractal
structure of the universe. 
This theory is powerfully predictive without any free parameter. 
It allowed to compute the $N$-points density correlations without any extra 
assumption\cite{gal}.

\section{Thermodynamics of the  Self-Gravitating Gas: the canonical
ensemble}

We investigate in this section  a  gas formed by  $N$ 
non-relativistic particles with mass $m$ interacting 
only through Newtonian gravity and which are in thermal
equilibrium at temperature $ T \equiv \beta^{-1} $.
We shall work in the  canonical ensemble assuming the gas being  on a
cubic box of side $ L $.

The  partition function of the system can be written as

\begin{equation}\label{fp}
{\cal Z} = \int\ldots \int
\prod_{l=1}^N\;{{d^3p_l\, d^3q_l}\over{(2\pi)^3}}\; e^{- \beta H_N}
\end{equation}
where
\begin{equation}\label{hamic}
H_N = \sum_{l=1}^N\;{{p_l^2}\over{2m}} - G \, m^2 \sum_{1\leq l < j\leq N}
{1 \over { |{\vec q}_l - {\vec q}_j|}}
\end{equation}
$G$ is Newton's gravitational constant.

Computing the integrals over the momenta $p_l, \; (1 \leq l \leq N) $

$$
\int\;{{d^3p}\over{(2\pi)^3}}\; e^{- {{\beta p^2}\over{2m}}} =
\left({m \over{2\pi \beta}}\right)^{3/2}
$$

yields

\begin{equation}\label{gfpc}
\displaystyle{
{\cal Z} =  \left({m \over{2\pi \beta}}\right)^{\frac{3N}2}
\; \int_0^L\ldots \int_0^L
\prod_{l=1}^N d^3q_l\;\; e^{ \beta G \, m^2 \sum_{1\leq l < j\leq N}
{1 \over { |{\vec q}_l - {\vec q}_j|}} }}
\end{equation}
We make now explicit the volume dependence introducing the new
variables $ {\vec r}_l ,\;  1\leq l \leq N $ as
\begin{eqnarray}
{\vec q}_l &=& L \; {\vec r}_l \quad , \quad {\vec r}_l =(x_l,y_l,z_l)
\;, \cr \cr
0&\leq& x_l,y_l,z_l \leq 1\; .
\end{eqnarray}
The partition function takes then the form,
\begin{equation}\label{fp2}
{\cal Z} = \left({m T L^2\over{2\pi}}\right)^{\frac{3N}2}
\; \int_0^1\ldots \int_0^1
\prod_{l=1}^N d^3r_l\;\; e^{ \eta \; u({\vec r}_1,\ldots,{\vec r}_N)}
\end{equation}
where we introduced the variable $ \eta $,
\begin{eqnarray}\label{defeta}
\eta &\equiv& {G \, m^2 N \over L \; T} \\ \cr \mbox{and}
\cr \cr
 u({\vec r}_1,\ldots,{\vec r}_N)   &\equiv& \frac1{N} \sum_{1\leq l < j\leq N}
{1 \over { |{\vec r}_l - {\vec r}_j|}} \nonumber
\end{eqnarray}
Recall that
\begin{equation}\label{Upot}
 U \equiv - {G \, m^2 N \over L}\; u({\vec r}_1,\ldots,{\vec r}_N) 
\end{equation}
is the potential energy of the gas.

The free energy  takes then the  form,
\begin{equation}\label{flib}
F = -T \log {\cal Z} = -3 N T \log\left(\sqrt{ {mT\over 2\pi}
}L\right) - T \; \Phi_N(\eta) 
\end{equation}
Here
\begin{equation}\label{FiN}
\Phi_N(\eta) = \log \int_0^1\ldots \int_0^1
\prod_{l=1}^N d^3r_l\;\; e^{ \eta \; u({\vec r}_1,\ldots,{\vec r}_N)}\; ,
\end{equation}
The derivative of the function $ \Phi_N(\eta) $ will  be computed by
Monte Carlo simulations and,  in the weak field limit $ \eta << 1 $,
it will be calculated analytically.

We get for the pressure of the gas,
\begin{equation}\label{pres}
p = - \left({ \partial F \over  \partial V}\right)_T = {N T \over V} -
{\eta \, T \over 3 \, V} \; \Phi_N'(\eta)\; .  
\end{equation}
[Here, $ V \equiv L^3 $ stands for the volume of the box].
We see from eq.(\ref{FiN}) that $ \Phi_N(\eta) $ increases with $ \eta
$. Therefore, the second term in eq.(\ref{pres}) is a {\bf negative}
correction to the perfect gas pressure $ {N T \over V} $. 

The mean value of the potential energy $ U $ can be written from
eq.(\ref{Upot}) as
\begin{equation}\label{umed}
<U> = - T \eta  \; \Phi_N'(\eta)
\end{equation}
Combining eqs.(\ref{pres}) and (\ref{umed}) yields the virial theorem,
$$
{p V \over N T} = 1 +{ <U>\over 3 N T}\; ,
$$
or more explicitly
\begin{equation} \label{estado}
{p V \over N T} = 1- {\eta \over 3 N} \; \Phi_N'(\eta)
\end{equation}
where,
\begin{eqnarray}
\Phi_N'(\eta) &=& e^{-\Phi_N(\eta)} \; \int_0^1\ldots \int_0^1
\prod_{l=1}^N d^3r_l\;  u({\vec r}_1,\ldots,{\vec r}_N)\; e^{ \eta
u({\vec r}_1,\ldots,{\vec r}_N)}\cr 
\cr  &=& \frac12(N-1)  \; e^{-\Phi_N(\eta)} \; \int_0^1\ldots \int_0^1
\prod_{l=1}^N d^3r_l\;{1 \over |{\vec r}_1-{\vec r}_2|}\; e^{ \eta
u({\vec r}_1,\ldots,{\vec r}_N)} 
\end{eqnarray}
This formula indicates that $ \Phi_N'(\eta) $ is of order $ N $ for
large $ N $. Monte Carlo simulations  as well as analytic calculations
for small $ \eta $  show that this is indeed the case. In conclusion,
we can write the equation of state of the self-gravitating gas as
\begin{equation}\label{pVnT}
{p V \over N T} = f(\eta) \quad ,     
\end{equation}
where the function 
$$ 
f(\eta) \equiv  1- {\eta \over 3 N} \; \Phi_N'(\eta)\; ,
$$ 
is {\bf independent} of $ N $  for large $N$ and fixed $ \eta $.
[In practice, Monte Carlo simulations show that $ f(\eta) $ is
independent of $ N $ for  $ N > 100 $]. 

We get in addition,
$$
<U>= -3 N T\;[ 1-  f(\eta)]\;.
$$

In the dilute  limit, $ \eta \to 0 $ and we find the perfect gas
value
$$
f(0) = 1\; .
$$

Equating eqs.(\ref{estado}) and (\ref{pVnT}) yields,
$$
\Phi_N(\eta)= 3N \, \int_0^{\eta} dx \, { 1 - f(x) \over x}\; .
$$

Relevant thermodynamic quantities can be expressed in terms of the
function $ f(\eta) $. We find for the free energy from
eq.(\ref{flib}),
\begin{equation}
F =  -3 N T \log\left(\sqrt{ {mT\over 2\pi}}\right) - 3NT\; \int_0^{\eta} dx
\, { 1 - f(x) \over x}\; . 
\end{equation}
We find for the total energy
$$
E = - 3NT[ \frac12 - f(\eta)]\;,
$$
for the chemical potential,
$$
\mu = \left({ \partial F \over  \partial N}\right)_{T,V} = -3T
\log\left(\sqrt{ {mT\over 2\pi}}\right) - 3T[1 - f(\eta)] - 3T\; \int_0^{\eta} dx
\, { 1 - f(x) \over x} 
$$
and for the entropy
\begin{eqnarray}\label{entro}
S &=& - \left({ \partial F \over  \partial T}\right)_V \cr \cr
&=& -\frac32 N + 3 N  \log\left(\sqrt{ {mT\over 2\pi}}\right)+ 3N\; \int_0^{\eta} dx
\, { 1 - f(x) \over x}+ 3 N \; f(\eta) \; .  
\end{eqnarray}

The specific heat at constant volume takes the form\cite{llms},
\begin{eqnarray}\label{ceV}
c_V &=& {T \over N}  \left({ \partial S \over  \partial T}\right)_V\cr \cr
&=& 3 \left[  f(\eta)-\eta \; f'(\eta) -\frac12 \right]\; .
\end{eqnarray}
where we used eq.(\ref{entro}).
This quantity is also related to the fluctuations of the potential
energy $(\Delta U)^2$ and it is positive defined,
$$
c_V =\frac32 + (\Delta U)^2 \; .
$$
Here,
\begin{equation}\label{delu}
(\Delta U)^2 \equiv {{<U^2>-<U>^2}\over N \; T^2}= 3 \left[
f(\eta)-\eta \; f'(\eta) - 1 \right]\; .
\end{equation}

The specific heat at constant pressure is given by \cite{llms}
$$
c_P = c_V -  {T \over N}{{ \left({ \partial p \over  \partial
T}\right)^2_V}\over { \left({ \partial p \over  \partial V}\right)_T}}\; .
$$

and then,
\begin{eqnarray}\label{ceP}
c_P &=& c_V + { \left[f(\eta)-\eta f'(\eta)\right]^2 \over
f(\eta)+\frac13 \eta f'(\eta)} \cr \cr
&=& -\frac32 + {{4\, f(\eta)\left(f(\eta)-\eta f'(\eta)\right)} \over
{f(\eta)+\frac13 \eta f'(\eta)}} \; .
\end{eqnarray}

The isotherm ($K_T$) and adiabatic ($K_S$) compressibilities take the form
$$
K_T = - { 1 \over V} \left({ \partial V \over  \partial p}\right)_T =
{V \over N\, T} {1 \over {f(\eta)+\frac13 \eta f'(\eta)}} \; ,
$$
$$
K_S = - { 1 \over V} \left({ \partial V \over  \partial p}\right)_S = 
{ c_V \over c_P} \; K_T \; .
$$ 

\bigskip

The speed of sound $ v_s $ can be written here as \cite{llmf}
$$
v_s^2 = - {{ c_P \; V^2} \over {c_V \; N}} \left({ \partial p \over
\partial V}\right)_T = {V^2 \over N} \left[ {T \over N \, c_V}  \left({
\partial p \over  \partial T}\right)^2_V  - \left({ \partial p \over
\partial V}\right)_T \right] \; .
$$
Therefore,
\begin{equation}\label{vson}
{v_s^2 \over T} = { \left[f(\eta)-\eta f'(\eta)\right]^2\over 3
\left[f(\eta)-\eta f'(\eta)-\frac12 \right]} + f(\eta)+\frac13 \eta
f'(\eta)\; .
\end{equation}
 
We see that the large $ N $ limit of the self-gravitating gas is
special. Energy, free energy  and entropy are extensive in the sense
that they are proportional to the number of particles $ N $ (for fixed $\eta$).
They all depend on the variable $ \eta ={G \, m^2 N \over L \; T} $
which is to be kept fixed for the thermodynamic limit
($ N\to \infty $ and $ V \to \infty $) to exist.
Notice that  $ \eta $ contains the
ratio $ N/L = N \; V^{-1/3} $ and it is not an intensive variable in
the usual sense. Here,  the presence of long-range gravitational
situations calls for a new type of  variables in the thermodynamic limit. 

\subsection{Short-distance cutoff}

At short distance the particle interaction for the self-gravitating
gas in physical situations is not gravitational. Its
exact nature depends on the problem under consideration (opacity limit,
Van der Waals forces for molecules etc.). 
We shall just assume a repulsive short distance potential. That is,
\begin{eqnarray}
v_a(r) &=& -{1 \over r} \quad \mbox{for} \; r \ge a \cr \cr
v_a(r) &=& +{1 \over a} \quad \mbox{for}\; r \le a 
\end{eqnarray}
where $ r \equiv |{\vec r}_l - {\vec r}_j| $ stands for the distance
between the particles and $ a << 1 $ is the short distance cut-off. 

The presence of the repulsive short-distance interaction prevents the
 collapse (here  unphysical) of the self-gravitating gas. In the situations we
are interested to describe (interstellar medium, galaxy distributions)
the collapse situation is unphysical. 

\subsection{The diluted regime: $ \eta << 1$}

We can obtain the thermodynamic quantities as a series  in powers
of $ \eta $ just expanding the exponent in the integrand of $
\Phi_N(\eta) $ [eq.(\ref{FiN})]. 

To first order in $ \eta $ we get,
\begin{eqnarray}\label{etach}
\Phi_N(\eta) &=& \eta\; \int_0^1\ldots \int_0^1
\prod_{l=1}^N d^3r_l\;\; u({\vec r}_1,\ldots,{\vec r}_N)  
+ {\cal O}(\eta^2)  \cr \cr
&=& \frac12 \, \eta \; (N-1) \int_0^1 \int_0^1 { {d^3 r_1 \; d^3 r_2} \over
{ |{\vec r}_1 - {\vec r}_2|}} + {\cal O}(a^2)
+ {\cal O}(\eta^2)  \cr \cr
&=& 3(N-1)\; b_0 \; \eta + {\cal O}(a^2)+ {\cal O}(\eta^2) \; .
\end{eqnarray}
where the coefficient $ b_0 $ is just a pure number. For the cubic
geometry chosen, it takes the value
$$
b_0 = \frac43 \;  \int_0^1 (1-x) \, dx \int_0^1 (1-y)\, dy\int_0^1 
{(1-z)\, dz \over \sqrt{x^2 + y^2 + z^2}} = 0.31372\ldots\; .
$$
To first order in $ \eta $ we see that the cutoff effect is
negligeable $ \sim {\cal O}(a^2) $.

We therefore find in the low density limit using eqs.(\ref{estado}),
(\ref{pVnT}) and (\ref{etach})
\begin{equation}\label{petach}
{p V \over N T} = f(\eta) = 1 -  b_0 \; \eta +  {\cal O}(\eta^2)\; ,
\end{equation}
in the large $ N $ limit.

Furthermore, the speed of sound approaches linearly in $ \eta $ to its
perfect gas value,
$$
{v_s^2 \over T} \; \;  {\buildrel{ \eta \downarrow 0}\over =}\; \;  \frac53
-\frac43\, b_0 \; \eta +  {\cal O}(\eta^2)\; .
$$
We used here eqs.(\ref{vson}) and (\ref{petach}).

\subsection{Monte Carlo calculations}

We have applied  the standard Metropolis algorithm to the
self-gravitating gas in a cube of size $ L $ at temperature $ T $. We
computed in this way the pressure, the potential energy fluctuations
and the average particle distance as functions of $ \eta $. 

Two different phases show up: for $ \eta < \eta_c $ we have a
non-perfect gas and for $  \eta > \eta_c $ it is  a condensed
system with {\bf negative} pressure. The transition between  the two phases
is very sharp. We can say that the phase transition  is of first order since
there is a jump in the entropy. However, it is an unusual phase
transition since the pressure is negative in the denser phase. In
particular, the phases cannot  coexist since the pressure has opposite
sign in the two phases. This
phase transition is associated with the Jeans unstability.

We plot in figs. 1-3 $ f(\eta) = pV/[NT] \; , (\Delta U)^2 $ and the
speed of sound  squared $ v_s^2 / T  $ as functions of $ \eta $.

We find that  for small $ \eta $,  the Monte Carlo results for  $ pV/[NT] $
well reproduce the analytical formula (\ref{petach}).  $ pV/[NT] $
monotonically 
decreases with  $ \eta $. When $ \eta $ gets close to the value   $
\eta_c \sim 1.6 $ a phase transition suddenly hapens and  $
pV/[NT] $ becomes large and negative. 

$ <r> $  monotonically decreases with  $ \eta $ too. Near  $ \eta_c \;
, <r> $ has a sharp decrease. 

In the Monte Carlo simulations the phase transition to the condensed
phase happens for $ \eta = \eta _T $ slightly below $ \eta_c $. For $
\eta _T < \eta < \eta_c $, the gaseous phase may only exist as a
metastable state. We find that $ \eta_c - \eta _T$ decreases with the
number $ N $ of particles in the simulation. For example, $ \eta_c -
\eta _T \sim 0.2 $ for $ N = 2000 $.

Both, the values of $ pV/[NT] $ and  $ <r> $ in the condensed phase
depend on the cutoff $ a $. The Monte Carlo results for $ \eta > \eta_c  $
can be approximated as 
$$
{pV \over NT} = f(\eta) \simeq 1 -  {\eta \over {K \; a}} \quad ,
\quad  <r> \simeq 1.5 \, a \; .
$$
where $ K \sim 30 $. Therefore, the latent heat of the transition is
$$
q \simeq T \; \left[1 -  {\eta \over {K \; a}}\right] < 0 \; .
$$
 The behaviour of $ pV/[NT] $ near  $ \eta_c $ in the gaseous phase 
can be well reproduced by 
\begin{equation}\label{pecrit}
{pV \over NT}= f(\eta) \;\; {\buildrel{ \eta \uparrow \eta_c}\over =}\;
 A \; (\eta_c-\eta)^B
\end{equation}
where $ A \simeq 0.68 , \eta_c \simeq 1.59\; $ and $ B \simeq
0.22 $.  

In addition, the
behaviour of $ (\Delta U)^2 $ in the same region is well reproduced by
\begin{equation}\label{flUc}
(\Delta U)^2\; \; {\buildrel{ \eta \uparrow \eta_c}\over =}\;\; C  \;
(\eta_c-\eta)^{B-1} 
\end{equation}
with $ C \simeq 0.2 $  and with the exponent $ B - 1 $ as it must be
since $(\Delta U)^2$ grows as $ f'(\eta) $ [see eq.(\ref{delu})]. 
[Notice that for finite $ N , \; (\Delta U)^2 $ will be finite albeit
very large at the phase transition]. 

We thus find a critical region just below $ \eta_c $ where the energy
fluctuations tend to infinity as $ \eta \uparrow \eta_c $.

\bigskip

$ v_s^2, \; c_V $ and $ K_S $ turn to be positive in the whole interval
$ 0 \leq \eta \leq \eta_c $ while $ c_P $ and $ K_T $ are positive
for $ 0 \leq \eta \leq \eta_0 $ and change sign 
(and diverge) at the point $ \eta_0 < \eta_c $ defined by
$$
 f(\eta_0)+\frac13 \; \eta_0 \; f'(\eta_0) = 0 \; .
$$
We find $ \eta_0 \simeq 1.49 $ from the Monte Carlo simulations.
We find that $ c_P $ and $ K_T $ are negative for $  \eta_0 < \eta < \eta_c $.

The specific heat behaviour near the transition follows from eqs.(\ref{ceV}),
(\ref{ceP}) and (\ref{pecrit}),
$$
c_P \; \; {\buildrel{ \eta \uparrow \eta_c}\over =}\;\; -\frac32 +
{\cal O}\left[(\eta_c-\eta)^B \right] \; ,
$$
$$
c_V \; \; {\buildrel{ \eta \uparrow \eta_c}\over =}\;\; 3 \, A \, B \,
\eta_c \, (\eta_c-\eta)^{B-1} -\frac32 +{\cal O}\left[(\eta_c-\eta)^B
\right] \; .
$$

That is, while $ c_P $ tends to a negative value, $ c_V $ grows
without bound when  $ \eta \uparrow \eta_c $.  The usual thermodynamic
inequalities\cite{llms}  forbiding such negative values do not apply here due
to the inhomogeneity of the self-gravitating gas at thermal equilibrium.

\bigskip

We find for the speed of sound near the phase transition from eqs.(\ref{vson})
and (\ref{pecrit}),
$$
{v_s^2 \over T}\;\;  {\buildrel{ \eta \uparrow \eta_c}\over =}\;\;
\frac16 + {\cal O}\left[(\eta_c-\eta)^B \right] \; .
$$
That is, the speed of sound tends to a constant value when $ \eta
\uparrow \eta_c $.  

In the condensed phase, $ v_s^2 $ becomes negative
indicating that there is no sound propagation in such state. 

We verified that the Monte Carlo results in the gaseous phase  
($ \eta < \eta_c $) are cutoff independent for $ 0.001 \geq a \geq 0.0 $. 

\subsection{Particle Distribution}

The particle distribution at thermal equilibrium obtained through the Monte 
Carlo simulations is inhomogenous both in the gaseous and condensed phases.


In the gaseous phase we find from the Monte Carlo values of the particle
density distribution that the mass $ {\cal M} $ within a volume $
{\cal V} = R^3 $ 
scales as
\begin{equation}
 {\cal M} = {\cal C} \; R^D
\end{equation}
where $ {\cal C} $ is a $R$ independent constant and $ D $ takes values
in the range,
$$
D = 1.9 - 2.2 \; .
$$
near the phase transition. That is for $ 1.4 \leq \eta < \eta_c $. For
smaller $ \eta $, $ D $ increases towards the value $ D = 3 $ for $
\eta \to 0 $. That is, the distribution becomes uniform in  the perfect gas
limit, as expected. 

The exponent $ D $ found here suggest the presence of a fractal
distribution near the  the critical point $ pV/[NT] = 0^+ $ for a very large 
number $N$ of particles. 

\section{Discussion}

We present here a set of new results for the self-gravitating thermal
gas obtained by Monte Carlo and analytic methods. In particular, they
confirm the general picture of the thermal self-gravitating gas.
Namely, a gaseous phase for higher
temperature and lower density and a condensed phase for lower
temperature and higher density\cite{sas,lynbell,pad}. Actually, we find
 more appropriate to characterize the phases by the sign of the pressure:
positive pressure in the gaseous phase and negative pressure in the
condensed phase. The pressure plays here the r\^ole of order parameter. 

The parameter $ \eta $ [introduced in eq.(\ref{defeta})] can be
related to the Jeans length of the system  
\begin{equation}\label{longJ}
d_J = \sqrt{3T \over m} { 1 \over \sqrt{G \, m \, \rho}} \; ,
\end{equation}
where $ \rho \equiv N/V $ stands for the number volume density. Combining
eqs.(\ref{defeta}) and (\ref{longJ}) yields
$$
\eta = 3 \left(L \over d_J \right)^2 \; .
$$
We see that the  phase transition takes place for $ d_J \sim L $. [The precise
numerical value of the proportionality coefficient depends on the geometry]. 
For $  d_J > L $ we find the gaseous phase and for $  d_J < L $ the
system condenses as   expected. 

Contrary to mean field treatments \cite{chandra,sas}, we {\bf do not
assume} here an equation of state but we {\bf obtain} the equation of state 
for the canonical ensemble [see eq.(\ref{pVnT})]. We find at the same time
that the relevant   variable is here $ \eta = G m^2 N/[V^{1/3}  T] $.
The relevance of the ratio $ G m^2 /[V^{1/3}  T] $ has been noticed on
dimensionality  grounds \cite{sas}. However, dimensionality arguments alone
cannot single out the crucial factor $ N $ in the variable $ \eta $.

The crucial point is that the thermodynamic limit exist if we
let  $ N \to \infty $ and $ V \to \infty $ {\bf keeping $ \eta $ fixed}. 
Notice that $ \eta $ 
contains the ratio $ N \; V^{-1/3} $ and not $ N / V $. This means that
in this thermodynamic limit $ V $ grows as $ N^3 $ and thus the volume density
$ \rho = N/ V $ decreases as $ \sim N^{-2} $. $ \eta $ is to be kept fixed
for a  thermodynamic limit to exist in the same way as the temperature.
 $ p V $, the energy $E$, the free energy, the entropy are functions of 
$ \eta $ and $ T $ times $N$. The chemical potential,
specific heat, etc. are just functions of $ \eta $ and $ T $.

\bigskip

The divergent growth of the energy fluctuations $ (\Delta U)^2 $ near
the phase transition has been previously noticed
\cite{sas,lynbell,pad}. We find here the precise behaviour of $
(\Delta U)^2 $ for $ \eta \uparrow \eta_c $ using Monte Carlo methods
[eq.(\ref{flUc})]. 

In refs.\cite{natu,prd,gal,eri} we worked in the grand canonical ensemble
around the point where  $ \log {\cal Z}_{GC} = 0 $. This  precisely corresponds
to   $  pV/[NT] = 0 $  which is the critical point in the
 canonical treatment given here. The presence of a critical
region where scaling holds supports the previous work in the
grand canonical ensemble \cite{natu,prd,gal,eri}.

\section{acknowledgements}

One of us (H J de V) thanks M. Picco for useful discussions on Monte
Carlo methods.

\begin{figure}[t] 
\epsfig{file=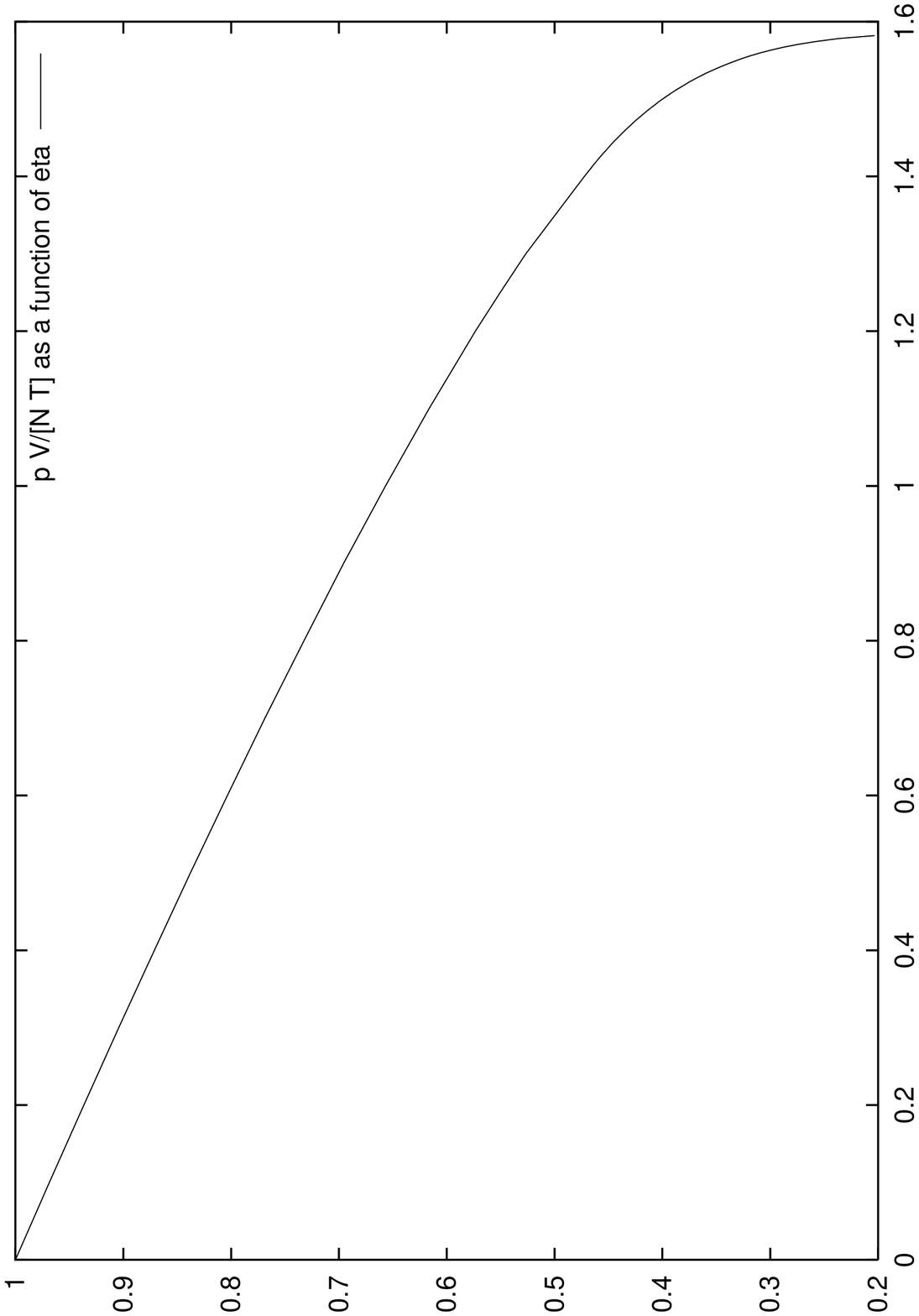,width=12cm,height=18cm} 
\caption{ $ f(\eta) = p V/[ N T] $ as a function of $ \eta $ in the gaseous
phase from Monte Carlo simulations. \label{fig1}} 
\end{figure}

\begin{figure}[t] 
\epsfig{file=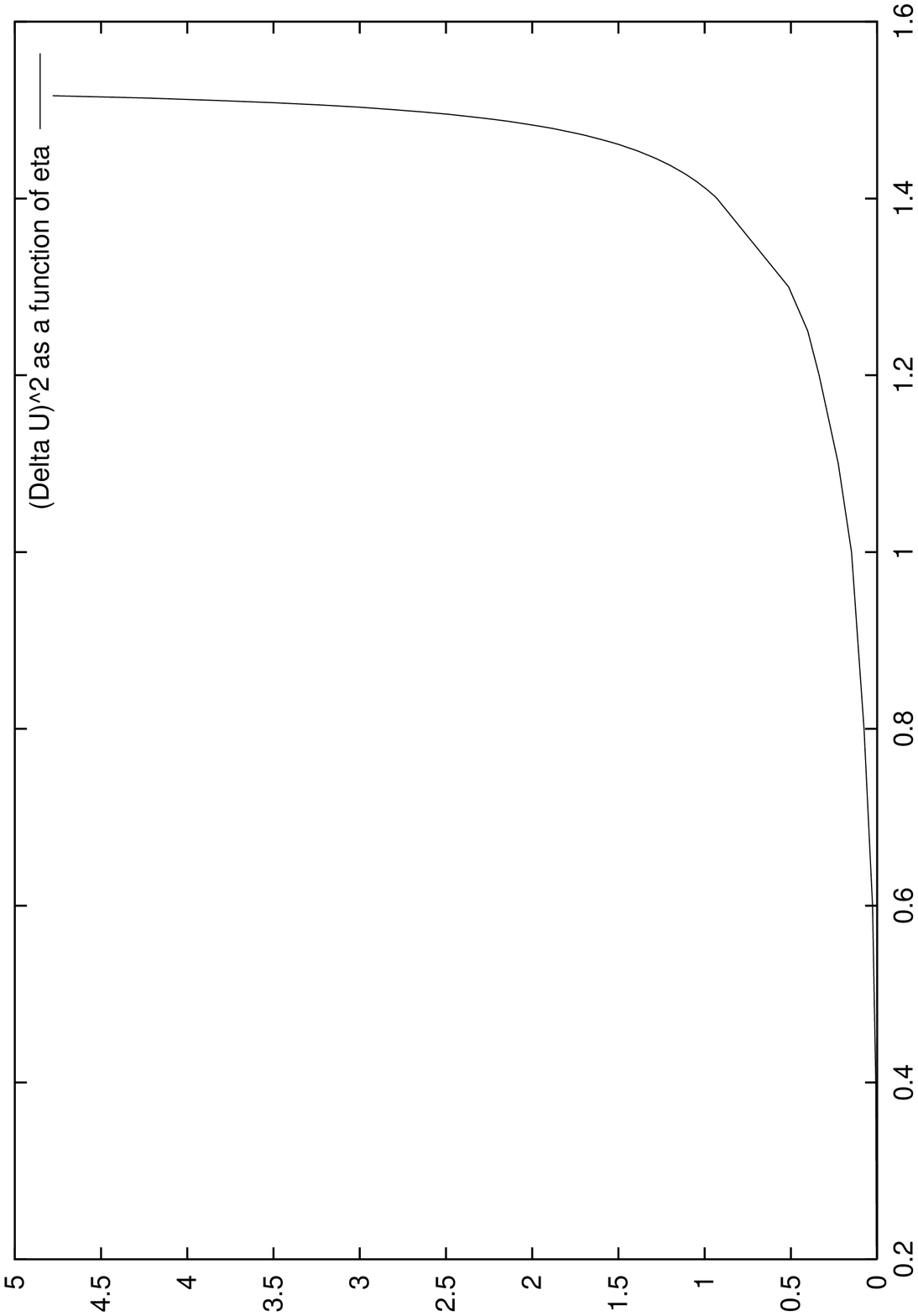,width=12cm,height=18cm} 
\caption{ $ (\Delta U)^2 \equiv {{<U^2>-<U>^2}\over N \; T^2}= 3 \left[
f(\eta)-\eta \; f'(\eta) - 1 \right]  $ as a function of $ \eta $ in
the gaseous phase from Monte Carlo simulations. Recall that $ c_V =
3/2 +  (\Delta U)^2 $. \label{fig2}} 
\end{figure}

\begin{figure}[t] 
\epsfig{file=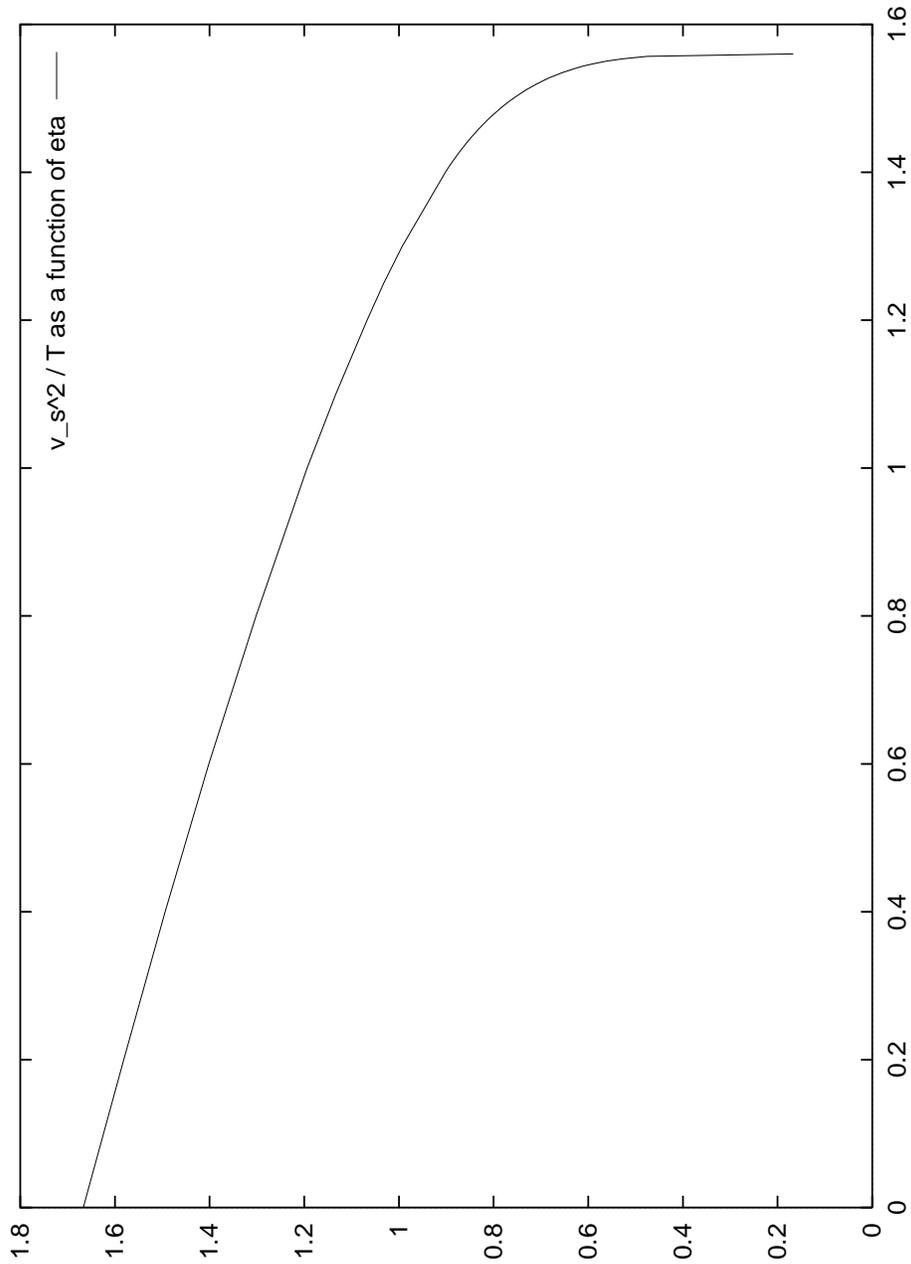,width=12cm,height=18cm} 
\caption{ The speed of sound squared divided by the temperature,
$ v_s^2 / T $,  as a function of $ \eta $ in
the gaseous phase from Monte Carlo simulations. Notice that $ v_s^2 /
T $ takes the value $ 1/6 $ at the critical point $\eta=\eta_c$.
\label{fig3}} 
\end{figure}

\end{document}